\documentclass[chicago]{emulateapj-rtx4}
\usepackage{amsmath,amssymb}


\shorttitle{MEMS Accelerometers}
\shortauthors{M\'esz\'aros et al.}

\begin{document}

\title{Accurate Telescope Mount Positioning with MEMS Accelerometers}


\author{L. M\'esz\'aros\altaffilmark{1},  
	A. Jask\'o, 
	A. P\'al\altaffilmark{1} and
	G. Cs\'ep\'any\altaffilmark{1,2}}
\affil{	MTA Research Centre for Astronomy and Earth Sciences, \\
        Konkoly Thege Mikl\'os \'ut 15-17,
        Budapest H-1121, Hungary}
\altaffiltext{1}{Department of Astronomy, Lor\'and E\"otv\"os University, 
		 P\'azm\'any P. stny. 1/A, 
		 Budapest H-1117, Hungary }
\altaffiltext{2}{ESO-Garching, Germany,
		D-85748, Karl-Schwarzschild-Str. 2}
\email{lmeszaros@flyseye.net, apal@flyseye.net}



\begin{abstract}
This paper describes the advantages and challenges 
of applying microelectromechanical
accelerometer systems (MEMS accelerometers) in order to attain precise, 
accurate and stateless positioning of telescope mounts. This provides
a completely independent method from other forms of electronic, 
optical, mechanical or magnetic feedback or real-time astrometry. 
Our goal is to reach the sub-arcminute range which is well smaller
than the field-of-view of conventional imaging telescope systems.
Here we present how this sub-arcminute accuracy can be achieved with 
very cheap MEMS sensors and we also detail how our procedures can be
extended in order to attain even finer measurements. In addition, our 
paper discusses how can a complete system design be implemented in order
to be a part of a telescope control system. 
\end{abstract}

\keywords{Techniques: photometric -- Instrumentation: miscellaneous}

\section{Introduction}

\noindent
The smooth and safe remote controlled, automatic or autonomous operation of 
telescopes are ensured via several independent and redundant mechanisms. Such 
mechanisms include electrical, mechanical, magnetic or optical limit switches, 
various forms of rotary or linear encoders, etc. The aim of this paper is to 
present an alternate approach for telescope mount position feedback by 
involving microelectromechanical accelerometer systems, also known as 
MEMS accelerometers \citep[for an introduction, see e.g.][]{lee2005,chollet2013}. 
These sensors are available in the
form of integrated circuits embedded in very compact packages. Such sensors
are capable to measure either static or dynamic acceleration where the
sensing directions and measurement ranges depend on the actual manufacturer 
and chip type. 

There are numerous ways to employ these MEMS accelerometers in 
telescope control systems (TCS). For instance, even a single channel 
accelerometer could act as a horizontal limit switch if its axis 
is mounted in parallel with the optical axis of the telescope tube
\citep[see e.g.][]{maureira2014}.
In the case of a telescope on an equatorial mount (which is 
located on temperate geographical latitudes), sensing the horizontal limit by 
measuring the two mount axes (hour angle and declination) implies
hardwired evaluation of trigonometric equations. In such a case,
microelectromechanical systems can be a viable alternative
in order to safely avoid the tube going below the horizon (or below some
another practical altitude). Accelerometers can, in addition, provide 
a location-independent way of handling such mount motion limitation.
Furthermore, telescope systems performing frequent re-positioning 
might require even more care in terms of redundancy and safe operations. 
Such systems include both survey instruments \citep[see e.g.][]{burd2005} 
and/or fast response devices \citep[see, for instance,][]{fors2013}.

Several pointing models exists for both equatorial 
\citep[see e.g.][]{spillar1993,buie2003} and alt-azimuth 
\citep[see e.g.][]{zhang2001,granzer2012} telescope mechanics. 
As we will see later on, alt-azimuth mounts cannot be used for a complete
pointing recovery by employing accelerometers. However, even equatorial
mechanics require a rather different approach during the interpretation
of the accelerometer outputs if such sensors are employed instead of 
more conventional solutions like rotary encoders. We have to note here
that geodesic and gravitational verticals differ due to local
anomalies \citep[see][]{hirt2006,hirt2008}, but such differences are
smaller by 2 orders of magnitude than our intended accuracy. 

The aim of this paper is to show how these integrated accelerometers
can be exploited in order to achieve an accurate mount positioning
at the level of an arcminute. In addition, the paper discusses the
aforementioned problem related to the interpretation of accelerometer
data regarding to pointing models. 
This paper is organized as follows. Sec.~\ref{sec:accelerometer} describes
the design of the hardware, firmware, software and data acquisition of
our set of accelerometer units. The calibration of the sensor is
performed in two major steps. First, Sec.~\ref{sec:calibration} describes
how the accelerometer itself can be calibrated, i.e. how can it act 
as a precise and accurate attitude sensor. Thereafter, 
Sec.~\ref{sec:pointing} details how can we accurately derive the 
telescope pointing from the output of the previously 
calibrated accelerometer, i.e. how can we calibrate the attitude of the
sensor itself with respect to the telescope mount components. 
Finally, in Sec.~\ref{sec:summary}
we summarize our work.

\begin{figure*}
\plottwo{th_dscf9308}{th_20131001920}
\noindent
\caption{
a) Left: one of the enclosed accelerometers as it is mounted on the center of
the fork of the hour axis mechanism of the Schmidt telescope. 
In this close-up view, one of the RJ45/8p8c plugs is connected.
b) Right: the accelerometer mounted on the telescope tube. The attitude
is rather arbitrary, the only constraint is that the optical axis
of the telescope lies in the accelerometer reference plane.}
\label{fig:accelerometers}
\end{figure*}

\begin{figure}
\plotone{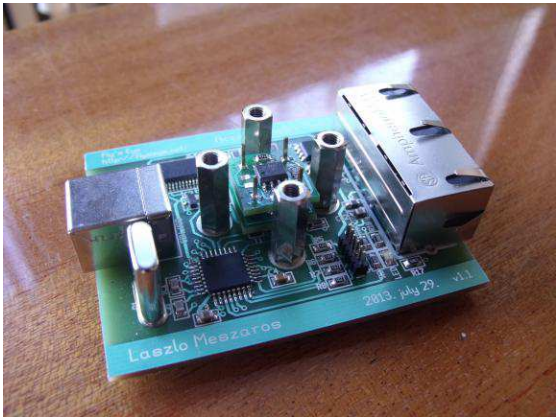}
\noindent
\caption{The accelerometer main electronics and the sensor boards
(mounted at the center of the main board). 
The left-side socket is an USB-B (``device side'')
while the right one is a dual RJ45/8p8c connector. The MCU is located
at the lower-left corner, next to the USB socket and the quartz.}
\label{fig:accelerometerboard}
\end{figure}

\section{The accelerometer design}
\label{sec:accelerometer}

\noindent
This section briefly describes the properties and features of the
complete accelerometer subsystem designed by our group. 
MEMS accelerometers usually packed as a 
surface mounted device (SMD) in a small form factor 
(usually quad-flat no-lead, QFN or leadless chip carrier, LCC)
and could provide analog, pulse-width
modulated (PWM) or completely digital interfaces, depending on the
manufacturer, the number of sensed axes (i.e. $\mathbf{g}$ vector components) 
and the actual type. However, independently from
the actual interface of these detectors, data rate is relatively high
and the output is undersampled. Namely, high data rate means approximately
kilosamples per second and the undersampled property yields quantized
Gaussian white noise output with a nearly unity standard deviation. 

Therefore, in order to both characterize the behaviour of
such sensors and employ these in an accurate TCS,
it is essential to build a higher level electronics, bus system and 
data acquisition (DAQ) frontend around the individual accelerometer chips. In the
following, we describe our solution to provide these features.
Some pictures of our assembly are exhibited in Fig.~\ref{fig:accelerometers}
and Fig.~\ref{fig:accelerometerboard}.

\subsection{Sensors and electronics}
\label{sec:electronics}

\noindent
Our choice for a MEMS accelerometer was the three-axis model MMA8453Q 
by Freescale, featuring a digital I$^2$C bus for complete data flow control
and some auxiliary bi-state output pins for other higher level applications
(e.g. landscape/portrait detection, free-fall detection, etc.). 
In our design, we exploited only the I$^2$C bus. Due to its intrinsic
properties, the accelerometer outputs depend on the temperature. In order
to compensate for the unexpected thermal responses, we employed two
high accuracy I$^2$C digital thermometers on two separate 
small ($12\,{\rm mm}\times12\,{\rm mm}$) 
circuit boards mounted below and above the accelerometer chip. These small
PCBs contain I$^2$C address selector resistors and bypass capacitors as well
and connected via each other and to the main board using 4 pins of 
ground, power ($+3.3\,{\rm V}$), SDA and SCL in the same arrangement and 
geometry as defined by the pins 1, 4, 5 and 8 of a DIP-8 package. In
fact, the main board contains one of the thermometers (the ``lower'' one)
while the daughterboards with the accelerometer chip and the other
(the ``upper'' one) thermometer are mounted above the main board
as it can be seen in Fig.~\ref{fig:accelerometerboard}. The whole
``building'' of these sensors is located at the geometric center of the
main board (within a precision of a tenth of a millimeter).

The I$^2$C bus master is the core MCU of the board, which is an AVR 8-bit 
microcontroller featuring 8\,kbytes of program space, 512\,bytes of static 
RAM and the similar amount of EEPROM. The program space is divided into a 
protected boot loader section and an application code section. Hence, the 
application code (the main firmware) can be upgraded easily via both kind of
serial interfaces (see later on in Sec.~\ref{sec:communication}). The
onboard electronics and firmware continuously poll the accelerometer
and thermometers and perform data binning in order to reduce the data flow
from $\approx$kilosamples per second down to approximately ten samples 
per second. The samples are queued in a dedicated memory area, 
hence bulk download of multiple binned data blocks are also possible
and (small) delays in the DAQ frontend even do not yield data loss. 
Furthermore, the binning procedure computes the standard deviations of the
individual measurements (on all of the axes) and provides these for the DAQ
controller. 

\begin{figure*}
\plottwo{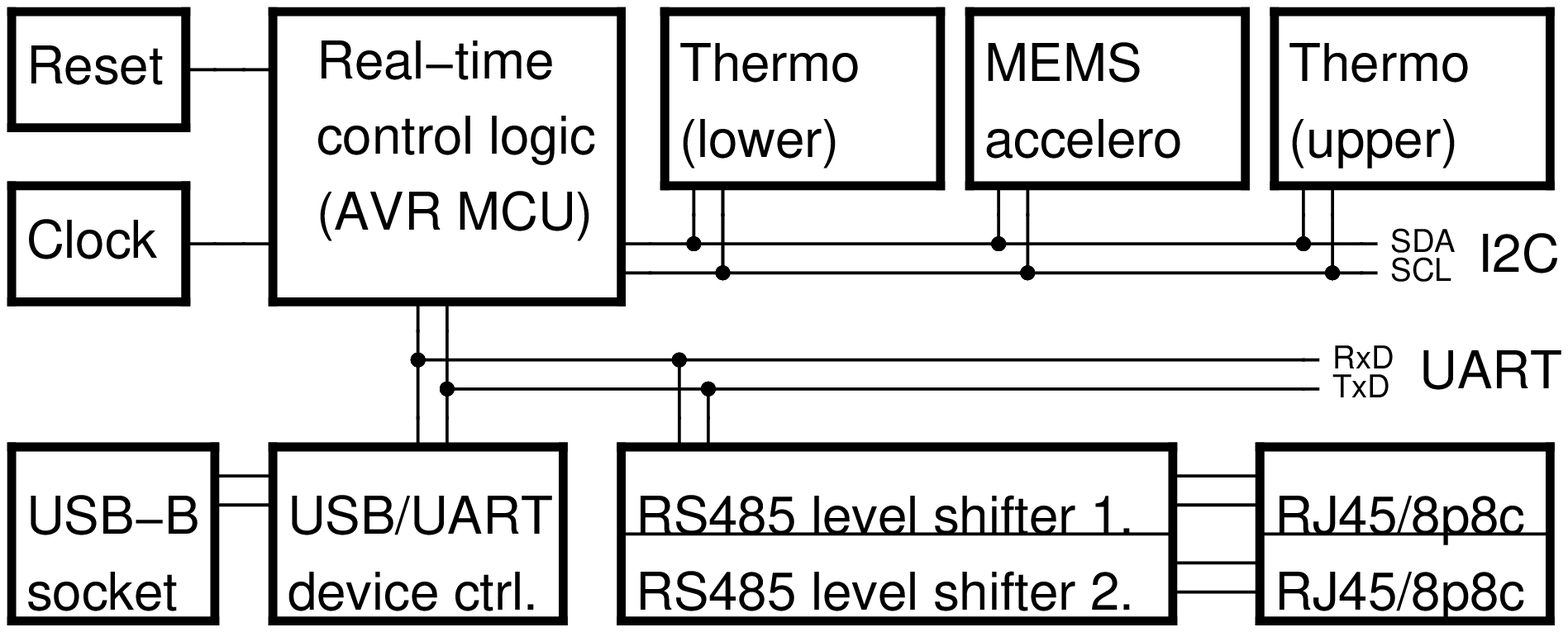}{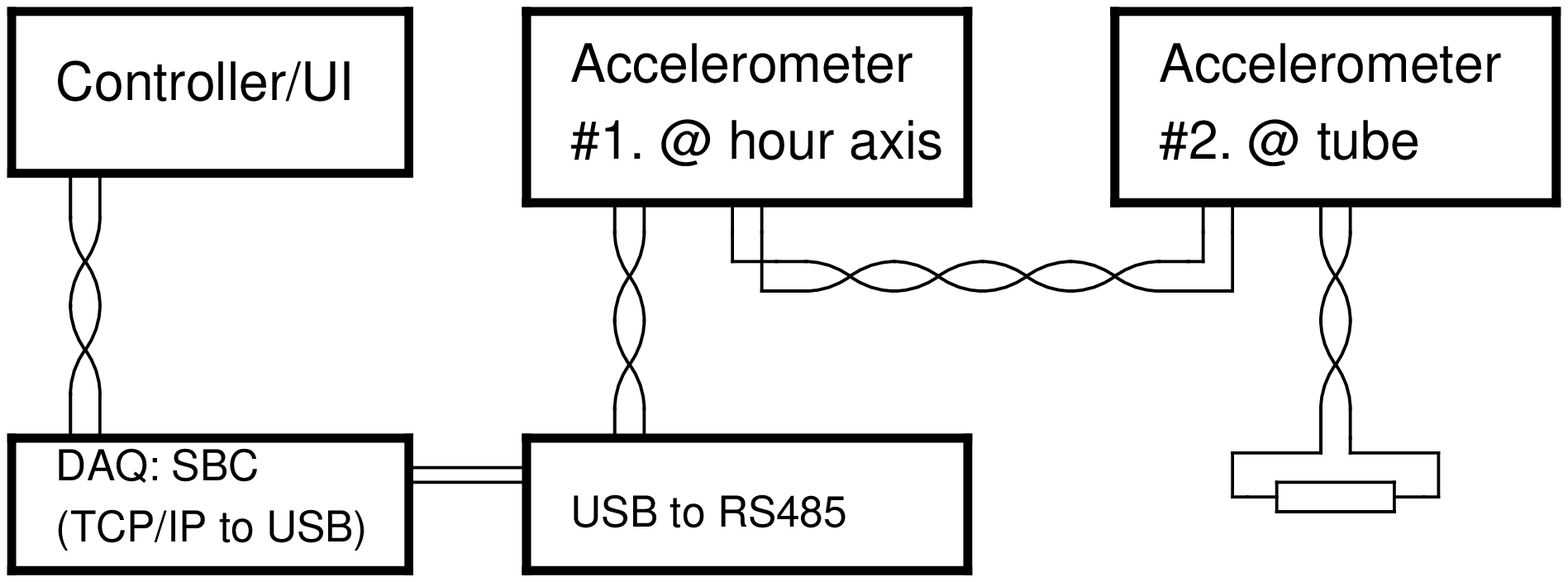}
\noindent
\caption{a) Left: block diagram of an individual accelerometer unit.
b) Right: block diagram of the complete subsystem with 
two accelerometers. In our test environment, the first one (\#1) is 
mounted on the hour axis (see also Fig.~\ref{fig:accelerometers}, center
and right panels) while the second one (\#2) is mounted on the telescope
tube itself.}
\label{fig:blockdiagram}
\end{figure*}

\begin{figure}
\plotone{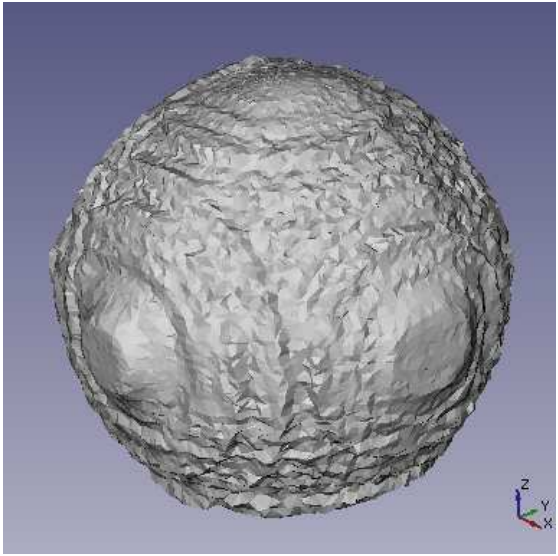}
\noindent
\caption{The residual of the spherically distributed points
after subtracting the best-fit affine transformation. The 
root mean square residual from the perfect sphere is $0.0021$. 
For clarity, in this image 
the residual itself is magnified by a factor of $10$. See text 
for further details.}
\label{fig:affineresidual}
\end{figure}

\begin{figure*}
\plottwo{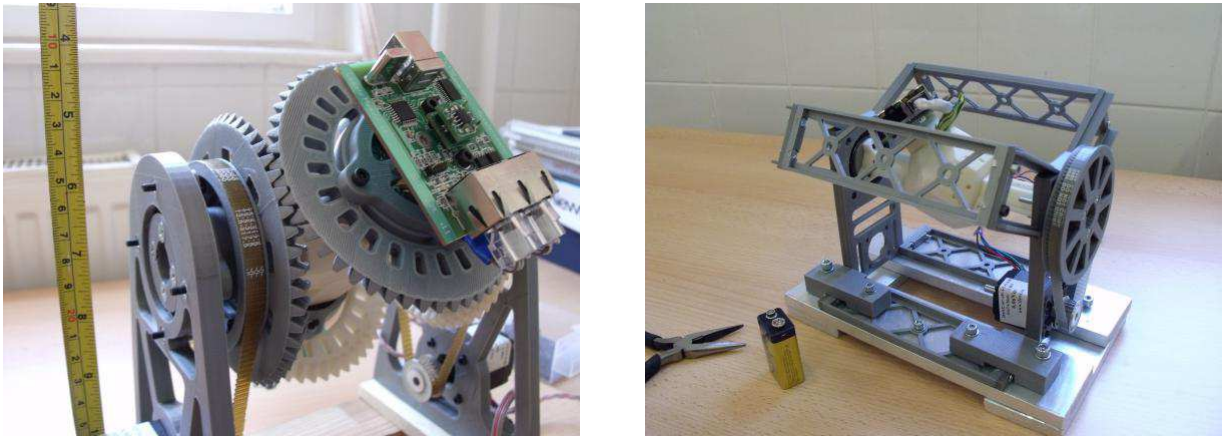}{th_dscf9558}
\noindent
\caption{Devices used
in our calibration procedure. a) Left: the two-axis setup used during
the analysis of the spherical constraints. The measuring tape shows the scale
in centimeters (left) and inches (right). b) Right: the single-axis setup 
used during the analysis of the planar constraints. The 9-volt (PP3) battery 
shows the scale.}
\label{fig:calibrationdevices}
\end{figure*}

\begin{figure}
\plotone{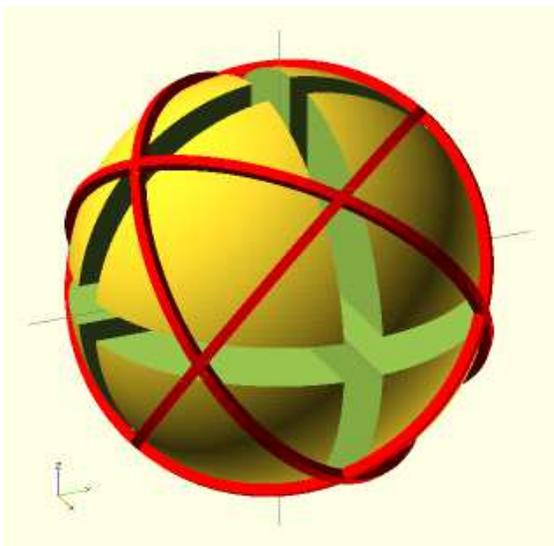}
\noindent
\caption{The unit sphere formed during the calibration procedure, 
showing both the sphere octants (yellow filled sections) and circles 
of the sphere that are nearly -- but not necessarily -- great 
circles (marked with dark red). For clarity, the distance $D_0$ 
specifying the gaps between the octants are drawn for $D_0=0.1$ in this
plot.}
\label{fig:calibrationsphere}
\end{figure}

\subsection{Bus system and communication protocol}
\label{sec:communication}

\noindent
The MCU hosts a single universal asynchronous receiver and
transmitter (UART) interface
that is currently connected to both a USB-UART device controller as well as 
to a dual RS485 level shifter. The RS485 bus I/O modes are controlled by 
the MCU while these three UART interfaces are multiplexed by 
simple boolean logic according to the UART standard. Hence, a USB host can be 
used for testing purposes or even short-distance DAQ while the
dual RS485 driver can be used to build a multi-drop serial network
of multiple accelerometers that are able to communicate either in 
half-duplex (RS485) or full-duplex (RS422) modes. The connector
of these RS485/RS422 interfaces are RJ45/8p8c sockets, wired in 
a similar fashion as defined by the 100M Power-over-Ethernet (PoE) standards.
Hence, normal out-of-the-box Ethernet cables can be used for connecting 
accelerometers as used to build wired local area networks (LANs).

The block diagram of the electronics related to a single accelerometer
unit -- including the sensors, I$^2$C bus (as described in 
the previous section) as well as these UART interfaces 
(detailed here) --  can be seen 
in Fig.~\ref{fig:blockdiagram}\,a. Our choice for an auxiliary 
USB interface was inspired by the fact that USB host controllers can be 
found on every present-day computer and it is not a kind of legacy interface
like RS232. Another advantage of USB is that it provides sufficient
power (unlike the RS232 that features only signal ground and control signals).
The board also includes a linear voltage regulator that provides the 
$+3.3\,{\rm V}$ supply of the sensor (see also Sec.~\ref{sec:electronicinterfaces}).

In order to ensure the safe data transmission on the RS485 bus between
multiple (daisy chained) accelerometer units, we employ a packet-oriented
master-slave 9-bit UART protocol for communication. The RS485 master
initiates the connection by addressing one of the accelerometers by 
its node identifier and sends a packet of 8-bit bytes that are multiplexed
with a 9th control bit. The command encoded in the packet 
implies whether an answer is expected or not (i.e. a multicast message
is not replied since the bus can only be driven by one unit). Hence, this
protocol allows us a packet-oriented interface that is rather simple
and its integrity can easily be traced. Since the USB and RS485/422 interfaces
are multiplexed, an 8-bit USB device controller must emulate the 9th bit
by appropriately setting the parity bits (for instance, mark and space
parities require less computation than even or odd parities, but the
former ones are not supported by all of the host and/or device controllers).

\subsection{Data acquisition and system setup}
\label{sec:daq}

\noindent
The USB-RS485 converter is connected to a single-board computer (SBC), on
which a TCP/IP server listens to packets, appropriately serializes
them to the RS485 bus and forwards the answer to the respective client. 
This TCP/IP client is the main DAQ frontend that can optionally be run 
on a different and/or remote PC. This client converts raw binary data to 
human-readable output. In our setup, it is possible both to run a 
single DAQ frontend that accesses multiple nodes in a round-robin fashion and 
to use two DAQ programs communicating only with a single accelerometer node. 
The block diagram of this setup is displayed in Fig.~\ref{fig:blockdiagram}\,b. 

\section{Calibration by constraints}
\label{sec:calibration}

\noindent
As it was mentioned in the Introduction, calibration of accelerometer
units are performed in two steps. The first step is performed 
independently from any other further knowledge related to the
intended application of the sensor itself. The second calibration
step is performed after mounting the sensor onto its targeted mechanism 
(e.g. a telescope tube or one of the telescope axes) and the goal is to 
derive the attitude of the sensor with respect to this particular mechanism. 
In this section, we detail the first step of the above described
two-step procedure while the second step is detailed 
in Sec.~\ref{sec:pointing}.
Throughout the next two subsections, specific values (noise magnitudes,
regression values, etc.) are correspond to one of the many accelerometer
units.

\subsection{Spherical constraints}
\label{sec:spherical}

\noindent
As it was detailed in Sec.~\ref{sec:accelerometer}, an accelerometer unit
delivers three raw coordinates that are the vector components of the 
acceleration with respect to the sensor. In the case of our application
where the accelerometer is mounted on a quasi-static mechanism,
this acceleration is equivalent to the standard local gravity. 
This assumption can safely be considered even if the telescope
performs smooth sidereal tracking. In this case, the ratio of the 
additional centrifugal acceleration and the standard gravity $g_0$ is 
going to be $L\Omega^2/g_0$, where $L$ is the characteristic size
of the instrument and $\Omega$ is the angular velocity of Earth rotation.
If $L$ is in the size of few meters, this ratio is going to be
smaller than $10^{-9}$, that is equivalent to $\approx 5$\,mas. 

In the previously discussed static (or quasi-static) configuration,
these three vector components $x$, $y$ and $z$ 
provided by the accelerometer unit should correspond to the relation
\begin{equation}
x^2+y^2+z^2=g_0^2.
\end{equation}
In practice, the sensors yield their output in dimensionless units
that are scaled to the standard gravity of Earth. Therefore, in the following
we will simply write this constraint in the form of 
\begin{equation}
x^2+y^2+z^2=1.\label{eq:unityxyz}
\end{equation}
Raw output from the three sensor channels do not comply 
with this relation due to systematic, random and quantization errors.
The magnitude of this deviation can be characterized easily by 
the scatter of the $r=\sqrt{x^2+y^2+z^2}$ values. For our choice of the sensor
the root mean square (RMS) of $r-1$ is $\approx 0.021$ if the $(x,y,z)$ values
are sampled nearly uniformly on the sphere. It can easily be examined
that if we add an uncorrelated Gaussian white noise of $\sigma$ 
to the outputs of a three-channel \emph{ideal} accelerometer, 
then the standard deviation
of the noisy $\sqrt{x^2+y^2+z^2}$ values is also $\sigma$. In general,
the question is how the values of $(x,y,z)$ have to be transformed
to $(x',y',z')$ in order to yield the smallest RMS for 
$\sqrt{x'^2+y'^2+z'^2}-1$.

Let us now consider a generic affine transformation $(x,y,z)\to(x',y',z')$
that has the form
\begin{equation}
\begin{pmatrix}x'\\y'\\z'\end{pmatrix}=
\begin{pmatrix}x\\y\\z\end{pmatrix}+
\begin{pmatrix}A_{xx}&A_{xy}&A_{xz}\\A_{xy}&A_{yy}&A_{yz}\\A_{xz}&A_{yz}&A_{zz}\end{pmatrix}
\begin{pmatrix}x\\y\\z\end{pmatrix}+
\begin{pmatrix}\Delta x\\\Delta y\\\Delta z\end{pmatrix}.\label{eq:affine}
\end{equation}
In this equation, there are $P=6+3$ unknowns: the $6$ components of the 
symmetric matrix $\mathbf{A}$ and the $3$ components
of the offset vector $(\Delta x,\Delta y,\Delta z)$. In order to obtain 
the best fit values of these $9$ unknowns that minimize the standard
deviations of $(x')^2+(y')^2+(z')^2$ from unity, let us consider 
the following procedure. For simplicity, let us denote
the accelerometer output vector by $\mathbf{r}=(x,y,z)$.
In general, such a transformation that is linear in its parameters like 
Eq.~(\ref{eq:affine}) can be written in the form
\begin{eqnarray}
x' & = & x+\sum_i p^x_if^x_i(\mathbf{r}), \label{eq:genx} \\
y' & = & y+\sum_j p^y_jf^y_j(\mathbf{r}), \label{eq:geny} \\
z' & = & z+\sum_k p^z_kf^z_k(\mathbf{r}). \label{eq:genz} 
\end{eqnarray}
Here the quantities $p^x_i$, $p^y_j$ and $p^z_k$ 
are the components of the parameter
vector (which has $9$ components in Eq.~\ref{eq:affine}).
Since the expected values for $p^{(x,y,z)}_i$ are in the range of $\sigma$,
a linear and iterative way can be constructed to figure out these values.
Let us sample the unit sphere in $N$ points from which a series
of $(x_\ell,y_\ell,z_\ell)$ vectors are known (where $1\le\ell\le N$).
The constraint that $(x'_\ell,y'_\ell,z'_\ell)$ has an unit length can then be 
reordered to have the form of
\begin{eqnarray}
\sum_i 2x_\ell p^x_if^x_i(\mathbf{r_\ell}) + \label{eq:reorder}
\sum_j 2y_\ell p^y_jf^y_j(\mathbf{r_\ell}) + \\
+ \sum_k 2z_\ell p^z_kf^z_k(\mathbf{r_\ell}) = 
1-(x_\ell^2+y_\ell^2+z_\ell^2)- \nonumber \\
-\left[\sum_i p^x_if^x_i(\mathbf{r_\ell})\right]^2 
-\left[\sum_j p^y_jf^y_j(\mathbf{r_\ell})\right]^2- \nonumber \\ 
-\left[\sum_k p^z_kf^z_k(\mathbf{r_\ell})\right]^2. \nonumber
\end{eqnarray}
The values of $1-(x_\ell^2+y_\ell^2+z_\ell^2)$ is in the 
range of $\sigma$ while the terms $[\sum_i(\dots)]^2$
are in the range of $\sigma^2$. Therefore, these latter three terms 
can be neglected in the first iteration. It can easily be recognized that
the remaining set of equations yield a linear least
squares problem for the values $(p^x_i,p^y_j,p^z_k)$ that can
be solved in a straightforward manner if the 
number of sampled points $N$ is larger than the number of parameters $P$.
In the following iteration, the values for $[\sum_i(\dots)]^2$
can be inserted from the results and the least squares optimization
is repeated by assuming these terms to be constants. This iteration
procedure is then repeated for a few times until convergence. 

At first, this procedure seems to be straightforward, however, many
questions arise.
\begin{itemize}
\item How can the sphere be mapped in $N$ points effectively and homogeneously?
The accelerometer device has to be rotated accordingly, then one has to 
wait a bit to settle the system (in order to
make the quasi-static assumption be valid), then read the output
of the accelerometer. 
\item What is the most suitable set of functions $(f^x_i,f^y_j,f^z_k)$
that can effectively be exploited in order to yield unity $(x')^2+(y')^2+(z')^2$
values? 
\item If one of the vector components, for instance, $x$ is relatively
small, then even a larger value for the respective $p^x_i$ component
perturb only slightly the value of $(x')^2$. Hence, such points
have smaller influence in the total least squares procedure. This property
implies an issue if the respective base function $f^x_i$ depends only 
on values of $x$ having small absolute values. 
\item Do the calibration results, i.e. the components 
of the $(p^x_i,p^y_j,p^z_k)$ parameter vector depend on the 
external environment? If so, how? 
\end{itemize}
In the following, we detail these problems in more details while
the last two issues are discussed in Sec.~\ref{sec:planar} and
Sec.~\ref{sec:environment}.

\begin{figure}
\plotone{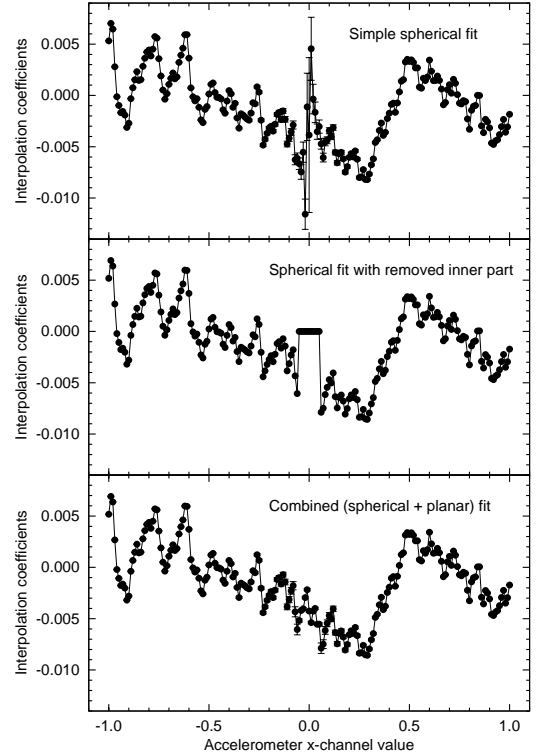}
\noindent
\caption{A typical reconstructed interpolation function for
the $x$-channel of one of our accelerometer units. The upper panel
shows the ``naive'' fit where only the spherical constraints were involved
in the reconstruction of the interpolation coefficients. It can 
be seen that for small $|x|$ values, the fit diverges
and the results become unreliable. The middle panel shows the results 
of the same fit while the values for $|x|\le D_0=0.05$ were forcibly set
to zero. The lower panel shows a completely reconstructed interpolation
function where spherical constraints were exploited for the two domains of
$x<-D_0$ and $D_0<x$ while planar constraints 
were used for the domain $|x|\le D_0$.}
\label{fig:interpolation}
\end{figure}

\subsubsection{A device for spherical mapping}
\label{sec:twoshaftdevice}

\noindent
In order to answer these questions and reflect to the problems mentioned above, 
first, we constructed a device that is capable to rotate the accelerometer
in a fashion that the accelerometer output vector moves on a spherical 
surface using a pre-defined pattern. The attitude of the accelerometer 
has three parameters, however, the sensor itself is not sensitive for 
the rotations around the vertical axis. Therefore, such a device can be
constructed using a combination of only two mechanisms where each of 
the mechanisms has one (rotational) degree of freedom. Likewise, it can be 
considered that if we are able to move a platform in a manner that
an arbitrary unit vector could completely scan the unit sphere,
then the output of an accelerometer fixed to this platform would also
completely scan the unit sphere. Our design for such a device is constructed
by involving parallel kinematics. Four bevel gears are connected in a 
similar fashion as it is used in a differential and two opposite gears
are driven separately by two motors. The cross connecting the four bevel
gears is able to freely rotate and the attitude of the two other bevel gears 
are determined by the rotational displacement of the two driven gears.
If the driven axis is horizontal, then the free gears will act as a suitable
platform that satisfies the previously discussed conditions. 

A photo from this device can be seen in Fig.~\ref{fig:calibrationdevices}\,a. 
This device is made using individually designed backlash-free bevel gear pairs.
These gears have been manufactured using 3D printing technology. 
The support structure, the differential cross and the bearing housings are 
also 3D printed parts. The cross and bearing shafts are hollow in order to 
easily connect the accelerometer with the data acquisition frontend and
to avoid unnecessary cable twisting. The two horizontal bevel gears are 
driven by timing pulleys. This solution both act as a reduction (therefore
allows a finer resolution) and lets the accelerometer cables freely 
leave the device via the hollow driven shaft. The timing pulleys are
driven by two stepper motors. The motor electronics share the same 
RS485 bus on which the accelerometer is 
connected (see Sec.~\ref{sec:communication}), hence a single program
can conduct the whole calibration procedure.

Since the duration of both the motor movements and the accelerometer 
data acquisition is in the range of a (few) second(s), 
several thousands of individual $(x_\ell,y_\ell,z_\ell)$
points can be retrieved in a few hours. Moreover, the device is capable
to support many accelerometers, thus the calibration procedure can be 
done in parallel for many units. The points of the 
sphere on which the accelerometer are sampled are on a nearly 
homogeneous triangle mesh, i.e. it forms a structure 
resembling a geodesic dome. 

\subsubsection{Regression functions}
\label{sec:sphericalregression}

\noindent
At the first glance, we employ the set of functions $(f^x_i,f^y_j,f^z_k)$
implied by the affine transformation of Eq.~(\ref{eq:affine}). 
This transformation needs $P=9$ parameters while for the least squares
fit, we involved $10{,}000$ individual data points. The white noise
component of each component of the acceleration vector was 
very close to $2\cdot 10^{-4}$. The fit yielded the values
\begin{eqnarray}
\Delta x & = & +0.020483 \pm 0.000018, \nonumber \\
\Delta y & = & -0.018311 \pm 0.000018, \nonumber \\
\Delta z & = & -0.000423 \pm 0.000018, \nonumber \\
A_{xx}	 & = & +0.006452 \pm 0.000026, \nonumber \\
A_{yy}	 & = & -0.003808 \pm 0.000026, \nonumber \\
A_{zz}	 & = & -0.006783 \pm 0.000025, \nonumber \\
A_{yz}	 & = & +0.001530 \pm 0.000020, \nonumber \\
A_{xz}	 & = & -0.000247 \pm 0.000020, \nonumber \\
A_{xy}	 & = & -0.000603 \pm 0.000020, \nonumber 
\end{eqnarray}
while the RMS of $\sqrt{x'^2+y'^2+z'^2}-1$ 
has been decreased to $0.0021$. This residual is 
significantly smaller than the raw residual by a factor of $10\times$.
This is pretty encouraging taking into account that the number of
degrees of freedom was $N-P=9{,}991 \gg P=9$ (in other words, the
fit results can easily be considered as unbiased). From the 
above list of fit parameters, one can easily deduce what are the 
characteristic values of the zero points and scalings of the
individual channels as well as the magnitude of crosstalks between
each channel. This latter quantity simply represent that 
the MEMS channels are not perpendicular to each other.

This residual of $0.0021$ is equivalent to $0.12^\circ=7.3^\prime$ angular
accuracy. However, it is still more than $10$ times
larger than the white noise
value of $2\cdot10^{-4}$, still indicating that this simple affine transformation
does not eliminate all of the systematic errors. Indeed, as it 
is shown in Fig.~\ref{fig:affineresidual}, these systematics can 
easily be recognized. 

In order to de-trend for these systematics errors, another assumption for
the $(x',y',z')\to(x'',y'',z'')$ transformation should be considered and applied
\emph{after} the evaluation of Eq.~(\ref{eq:affine}). 
Expecting that after subtracting the affine part
which is responsible for the cross-talk between the axes, the 
transformation can be separated to some $x'\to x''$, $y'\to y''$ and $z'\to z''$
functions. In the following, we search these functions in a form of a 
piecewise linear functions that are tabulated using an equidistant manner
in the interval $[-1,1]$. Let us consider a spacing of $\Delta = 1/N_{\rm inter}$
between these interpolation points. If one includes the boundaries
($\pm1$), then $2N_{\rm inter}$ intervals and $2N_{\rm inter}+1$ control points
are given. The interpolation function is then characterized by 
$3\times(2N_{\rm inter}+1)$ unknowns for all of the three axes. The piecewise
linear interpolation is then written in the form
\begin{eqnarray}
x'' & = & x'+C^{(x)}_{L(x')} \left[R(x')-\frac{x'}{\Delta}\right]+
C^{(x)}_{R(x')} \left[\frac{x'}{\Delta}-L(x')\right], \label{eq:intx} \\
y'' & = & y'+C^{(y)}_{L(y')} \left[R(y')-\frac{y'}{\Delta}\right]+
C^{(y)}_{R(y')} \left[\frac{y'}{\Delta}-L(y')\right], \label{eq:inty}  \\
z'' & = & z'+C^{(z)}_{L(z')} \left[R(z')-\frac{z'}{\Delta}\right]+
C^{(z)}_{R(z')} \left[\frac{z'}{\Delta}-L(z')\right], \label{eq:intz} 
\end{eqnarray}
where $L(\cdot)$ and $R(\cdot)$ are integers and defined as 
\begin{eqnarray}
L(t) & = & \left\lfloor\frac{t}{\Delta}\right\rfloor, \\
R(t) & = & L(t)+1. 
\end{eqnarray}
Here $\lfloor\cdot\rfloor$ denotes the floor function.
For simplicity, the interpolation coefficients $C^{(\cdot)}_m$ 
are indexed between $-N_{\rm inter} \le m \le N_{\rm inter}$.
As the residual after the affine transformation was $0.0021$ (see earlier),
we expect that the magnitudes of the coefficients $C^{(\cdot)}_m$
are also within the range of $|C^{(\cdot)}_m|\lesssim (1.5\dots 2.5)\times 0.0021$.
It can be seen that
Eqs.~(\ref{eq:intx}), (\ref{eq:inty}) and (\ref{eq:intz})
are merely special cases of 
Eqs.~(\ref{eq:genx}), (\ref{eq:geny}) and (\ref{eq:genz}). Hence, we can
apply Eq.~(\ref{eq:reorder}) in order to recover the coefficients
$C^{(\cdot)}_m$ in a similar least squares fashion as it was performed
in the case of the affine transformation. Since the respective
$p_i^{x,y,z}$ values are smaller by an order of magnitude than in the
affine case (Eq.~\ref{eq:affine}), the number of iterations are also
smaller. 

The number of points that are needed for the interpolation depends 
on the nature of the residual structure. We found that this specific
MEMS accelerometer chip built into our sensors needs $2N_{\rm inter}=200$ 
interpolation intervals for a viable reconstruction. However, it 
should be kept in mind that the total number of interpolation control
points, $3\times(2N_{\rm inter}+1)$ should not exceed the total number
of points sampled on the sphere, $N$. 

As it was noted earlier in this section, points with small $x$, $y$ or $z$
values could cause trouble since the square of these values
yields only a small increment in the value of $x^2+y^2+z^2$. Indeed,
the upper panel of Fig.~\ref{fig:interpolation} shows that
the conditions used in our regression (i.e. $N=10,000$, $N_{\rm inter}=100$
and a typical white noise of $2\cdot10^{-4}$) results in an interpolation
function that is ``unstable'' around $|x|, |y|, |z| \lesssim D_0=0.05$.
Therefore, we should add additional constraints to the whole procedure
in order to have a reliable fit in the complete $(x,y,z)\in[-1,1]$ domain.
In the following, we describe a possible method for such a reconstruction.

\subsection{Planar constraints}
\label{sec:planar}

\noindent
As it was concluded at the end of Sec.~\ref{sec:spherical}, that pure
spherical constraints are inadequate for the calibration of the
accelerometer in the domain of $|x|, |y|, |z| \lesssim D_0$. It can 
easily be seen that the portion of the spherical surface that is
affected by this effect is roughly $3D_0=15\%$ in total. This is a quite large area
that cannot be neglected. 

\subsubsection{A device for planar mapping}
\label{sec:singleshaftdevice}

\noindent
In order to resolve this problem, we designed and built an additional 
calibration device that aids the calibration on this domain. This
device has a single, nearly horizontal shaft on which the accelerometer
itself is mounted. This horizontal axis is rotated in small steps
and the accelerometer channels are read accordingly. The accelerometer
has a specific attitude with respect to this horizontal shaft.
During rotation, the output vectors form a circle which deflects roughly 
equally from all of the three axes. In other words, the normal vector 
of this circle is close to $(\pm 1/\sqrt{3},\pm 1/\sqrt{3},\pm 1/\sqrt{3})$. 
It can easily be considered that there are four possible attitudes of the
accelerometer unit with respect to the shaft that yields such a 
configuration in the resulting circles. Fig.~\ref{fig:calibrationsphere}
displays these four circles (as well as the domain on which the purely
spherical constraints described earlier are viable). It can also
be recognized that if one of the axes, for instance, the values of
$x$-channel are close to zero, then the other two axes have a 
value of $\approx \pm1/\sqrt{2}$. Since the intersection of a spherical
surface and a plane always produce a circle, the constraint that
is created by this single shaft device can be coined as a planar constraint.

Our device that performs this rotation around a single, nearly horizontal
shaft is exhibited on the right panel of Fig.~\ref{fig:calibrationdevices}.
Similarly to the four-geared mechanism, this device also features a hollow
shaft driven by a timing pulley. Hence, wiring is quite easy in this 
case as well. At the center of the shaft, a special polyhedral structure
is installed. The role of this part is to simply attach even more
(currently, up to four) accelerometer units and to attain
the previously noted four possible attitudes without too much effort. 

As we will see later on it is not essential to have a perfectly 
horizontal shaft around which the accelerometer is rotated. However,
the angle between the vertical and this axis must not
alter during a measurement cycle. This stability
is needed only during a single run: while swapping between the
four possible attitudes, one can alter the attitude of the shaft with respect
to the vertical as well. 

\subsubsection{Regression functions}
\label{sec:planarregression}

\noindent
As it was noted earlier, the circle measured by the accelerometer 
(during the rotation of the horizontal axis) is the intersection
of the sphere and an appropriate plane. This plane can be characterized
by the equation
\begin{equation}
n_xx+n_yy+n_zz=C. \label{eq:planeequation}
\end{equation}
This equation has four parameters: the three components 
of the plane normal $(n_x,n_y,n_z)$ as well as the constant $C$. 
However, this equation is homogeneous: the implied ambiguity between these
four parameters can be resolved by applying the constraint
\begin{equation}
n_x^2+n_y^2+n_z^2=1. \label{eq:normalunity}
\end{equation}
In other words, the normal vector $(n_x,n_y,n_z)$ 
should have unity length. As it was discussed 
above, the single shaft device maps these planes to be
\begin{equation}
|n_x|,|n_y|,|n_z|\approx 1/\sqrt{3}. \label{eq:nxyznorm}
\end{equation}
If Eq.~(\ref{eq:normalunity}) is satisfied, $C$ is going to 
be the cosine of the angle between the shaft and the vertical. 
The radius of the circle (i.e. the intersection of the unit sphere 
and this plane) is then $\sqrt{1-C^2}$.

The calibration procedure, i.e. the reconstruction of the $C^{(\cdot)}_m$
constants for $|m| \le D_0/\Delta$ is performed as follows. 

First, using the points for which $D_0/\Delta<|m|$, we
de-trend the accelerometer outputs using the previously obtained affine
coefficients followed by the interpolation procedure. These
points are the parts of the red circles in Fig.~{\ref{fig:calibrationsphere}
that lie on the yellow filled octants.

Second, these de-trended points are then substituted into 
Eq.~(\ref{eq:planeequation}) and by employing a least-squares fit
and the additional constraint defined by Eq.~(\ref{eq:normalunity}),
the values of $n_x$, $n_y$, $n_z$ and $C$ are computed. We note here that
the aforementioned stability of the angle between the rotation axis
and the vertical (see also at the end of Sec.~\ref{sec:singleshaftdevice})
can easily be quantified by the residual of this fit. If this fit
yields a residual that is significantly larger than the residual obtained
during the fit of the piecewise linear interpolation coefficients
in the domain of $D_0/\Delta<|m|$, then the respective 
measurement has to be repeated. 

In the final, third step, the constants $C^{(\cdot)}_m$ where $|m| \le D_0/\Delta$ 
are fitted via a simple linear least squares manner by
minimizing the merit function
\begin{eqnarray}
\chi^2 & = & 	\sum\limits_{|x| \le D_0}\left[(n_xx'+n_yy'+n_zz')-C\right]^2+ \label{eq:chi2planar}\\
& &		\sum\limits_{|y| \le D_0}\left[(n_xx'+n_yy'+n_zz')-C\right]^2+\nonumber\\
& &		\sum\limits_{|z| \le D_0}\left[(n_xx'+n_yy'+n_zz')-C\right]^2 \nonumber.
\end{eqnarray}
where the values for $x'$, $y'$ and $z'$ are given by 
Eqs.~(\ref{eq:intx}), (\ref{eq:inty}) and (\ref{eq:intz}), respectively.
By cause of Eq.~(\ref{eq:nxyznorm}) and $D_0\lesssim 0.05$, 
the three conditions appearing in the summations of Eq.~(\ref{eq:chi2planar})
are disjoint.

As it can be seen in the lower panel of Fig.~\ref{fig:interpolation},
this procedure is capable to provide reliable values for the 
domains where $|x|$, $|y|$ or $|z|$ are smaller than this limit of $D_0$.
The practical choice for $D_0$ depends on the actual $S/N$ values for
the detector, the number of points sampled in the sphere and the number
of points sampled in these planes/circles. 
\vspace*{3mm}

By combining the aforementioned two methods, i.e.
spherical and planar constraints, the RMS
residual from the perfect unit sphere is going to be in the range of
$2.3 \dots 2.6\times10^{-4}$ for our accelerometer sensor units. This residual is 
equivalent to $0.013\dots0.015^\circ\approx 0.8\dots0.9^\prime\approx 48\dots54^{\prime\prime}$.
This is only slightly larger than the white noise component of the individual
components, meaning that the calibration procedure yields an accuracy 
comparable to the random noise. \vspace*{5mm}

\subsection{External environment}
\label{sec:environment}

\noindent
In the following, we discuss how the foregoing procedures depend on
the external environment. We focus on the effects of variations
in the ambient temperature and the local gravity of the place where 
the calibration procedure takes place. 

\subsubsection{Ambient temperature}
\label{sec:ambient}

\noindent
Accelerometers are also sensitive to the variations in the ambient temperature. 
This is an intrinsic property of the capacitive moving part of the sensor 
system, and not only resulted by the semiconductors of the integrated 
electronics, see also \cite{dai2010}.
As we introduced
earlier, our accelerometer units incorporate two precise digital thermometers
mounted close to the MEMS chips. The two thermometers are mounted right
above and below the accelerometer detector in a symmetric arrangement.
Therefore, polling these sensors can provide reliable information not
only about the accelerometer temperature but the thermal gradient 
in its vicinity. 

Throughout the calibration procedure in our experiments, the ambient 
temperature was not controlled actively but the environment was quite settled.
During the $\approx4$\,hours of this procedure, the measured temperature
was $22.86\pm0.16^\circ{\rm C}$ (RMS) while the gradient between the 
two thermometers was $-0.11\pm0.02^\circ{\rm C}$ (RMS). Based on the
datasheets, this difference is definitely smaller than the accuracy of each 
sensor (which is actually $0.3^\circ{\rm C}$), therefore we can safely consider
isothermal conditions within the sensor package. All of the values 
(residuals, plots, etc.) presented earlier in this section were based 
on the measurements acquired in such circumstances.

Subsequently, the spherical constraint measurements has been repeated 
for $N=1{,}000$ points in a colder environment, namely 
$8.88\pm0.14^\circ{\rm C}$. This is a slight ($\Delta T=-13.98^\circ{\rm C}$) 
but significant difference in the ambient temperature. If the data series
from this colder measurements are de-trended using the best-fit data based
on the warmer series (see earlier), the residual from the perfect sphere
goes up to $\approx 0.0019$. This RMS value implies a thermal dependence 
of $\approx 13\times 10^{-5}/{\rm K}$ in the accelerometer channel outputs.
Using this de-trended series, we applied the procedure described in
Sec.~\ref{sec:sphericalregression}. This affine fit yielded a residual
of $0.00021$ which is in the range of the residual after the 
interpolation-based regression in the case of the warm data series.

Hence, we can conclude that the thermal dependence of the accelerometer
outputs can be obtained in two steps. First, one applies a full
(affine, spherical interpolation and planar interpolation) fit for a
certain ambient temperature $T_0$. Then, using these fit parameters, spherical
measurements gathered on a different ambient temperature $T_c$
are de-trended and this output are fitted again but only for the 
affine coefficients. Let us denote these affine coefficients 
by $\Delta\hat x$, $\Delta\hat y$, $\Delta\hat z$, $\hat{A}_{xx}$, 
$\hat{A}_{yy}$, \dots.
If the raw values $(x,y,z)$ are read at temperature $T$, then
in first step we apply Eqs.~(\ref{eq:affine}), (\ref{eq:intx}), 
(\ref{eq:inty}) and (\ref{eq:intz}) using the coefficients obtained
at $T_0$. Next, in the second step we apply Eq.~(\ref{eq:affine}) 
using the affine coefficients 
$k\Delta\hat x$, $k\Delta\hat y$, $k\Delta\hat z$, $k\hat{A}_{xx}$,
$k\hat{A}_{yy}$, \dots. Here $\Delta\hat x$, $\Delta\hat y$, etc. 
are obtained at the ambient temperature $T_c$ (see above) and 
\begin{equation}
k=\frac{T-T_0}{T_c-T_0}.
\end{equation}
This linear temperature dependence can be characterized
more accurately by taking further measurements on various other
ambient temperatures and/or by increasing the difference between $T_0$
and $T_c$. However, such a linear approximation can be feasible 
on even larger temperature ranges \citep[see e.g.][]{dai2010}.

\subsubsection{Electronic interfaces}
\label{sec:electronicinterfaces}

\noindent
The responses of analog circuits (including MEMS accelerometers)  
depend on the voltage levels -- most prominently, the power supply --
applied to these electronics. In order to ensure the stability and accuracy
of the whole sensor system, the supply voltage of the sensor should also be stabilized.
For this purpose, we employed an onboard linear regulator 
that provides the nominal $+3.3\,{\rm V}$ supply of the sensor and this
voltage is derived from the bus power. Such linear regulators safely
reduce the relatively large variations that are allowed by, e.g.
the USB standard (where it is $5.00\pm0.25\,{\rm V}$).
In addition, care must be taken in order to remove high-frequency components
appearing on the bus(es). Hence, bypass capacitors are included in both 
the bus side ($+5\,{\rm V}$) and sensor side ($+3.3\,{\rm V}$) of the circuit.

\subsubsection{Local gravity}
\label{sec:gravity}

\noindent
If the local gravity changes throughout the calibration procedure,
then the assumption of Eq.~(\ref{eq:unityxyz}) won't be true anymore.
The relative change of the local gravity depends on the location
of the Earth (i.e. the gravity itself is larger at the poles and 
smaller close to the equator) as well as it depends on the altitude.
The altitude dependence of $g_0$ can be characterized as
\begin{equation}
\frac{\Delta g_0}{g_0}=-\frac{2\Delta h}{R_0},\vspace*{0mm}
\end{equation}
where $\Delta h$ is the change in the altitude and $R_0$ is the radius
of the Earth. For instance, going up by $100\,{\rm m}$ yields a 
decrease of $\Delta g_0/g_0\approx -3\cdot10^{-5}$ while going north
by $100\,{\rm km}$ yields an increase of $\Delta g_0/g_0\approx +5\cdot10^{-5}$
on average, due to the oblate shape of the Earth. Furthermore,
the RMS residual from the reference ellipsoid is in the range of 
$\sigma(\Delta g_0/g_0) \approx 2\cdot10^{-5}$. The magnitude of
these effects are bit smaller but comparable to the residual of the 
accelerometer calibration procedure (see above at the 
end of Sec.~\ref{sec:planar}). Consequently, such effects must be
taken into account during the relocation of a calibrated device. 

\begin{figure}
\plotone{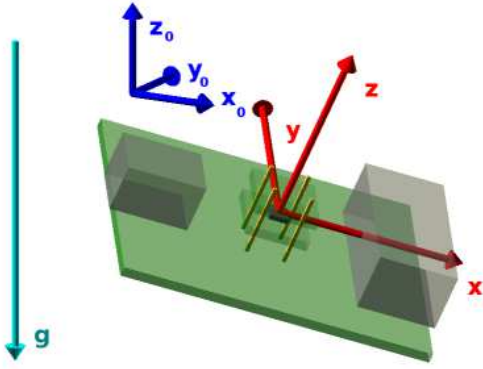}
\noindent
\caption{The coordinate system with respect to the accelerometer,
as represented with the red $(x,y,z)$ axes and the reference frame
of the environment, marked with the blue $(x_0,y_0,z_0)$ axes. 
The local gravity $\mathbf{g}$ points towards to the $z_0-$ direction. 
Although the geodesic and gravitational verticals differ, this
difference is significantly smaller than our intended accuracy.}
\label{fig:coord}
\end{figure}

\section{Pointing models}
\label{sec:pointing}

\noindent
As we emphasized in the introduction, static accelerometers are not
sensitive for rotation around the vertical axis as well as these do 
not sense displacements. The \emph{information} provided by a single
static accelerometer is a vector with unity length, i.e. a point on the
surface of a sphere. 

In order to examine various properties of the accelerometer units
while these are act as a telescope pointing sensor,
we installed two units to the Schmidt telescope of the Konkoly Observatory,
located at the Piszk\'estet\H{o} Mountain station. One of the
accelerometers is mounted on the fork of the telescope (see also
Fig.~\ref{fig:accelerometers}\,a) while the 
other one is mounted on the side of the telescope tube
(Fig.~\ref{fig:accelerometers}\,b). Both accelerometers
are fixed in a kind of arbitrary attitude, the only constraint is
that the $z+$ axis of the first unit sensor (\#1, mounted on the fork) is
roughly parallel with the hour axis and points towards north
while the $z+$ axis of the second sensor (\#2) is more-or-less perpendicular 
to the optical axis. For this
setup, we used a USB-RS485 converter as a bus master which is connected to
unit \#1, followed by unit \#2. The termination resistors of the bus are
placed after unit \#2. 

In this section, we investigate how can one derive the output of these
accelerometers. The following series of computations does not
depend on our actual setup, it can simply be adopted to any equatorial
telescope mount. 

\begin{figure*}
\begin{center}
\includegraphics[height=55mm]{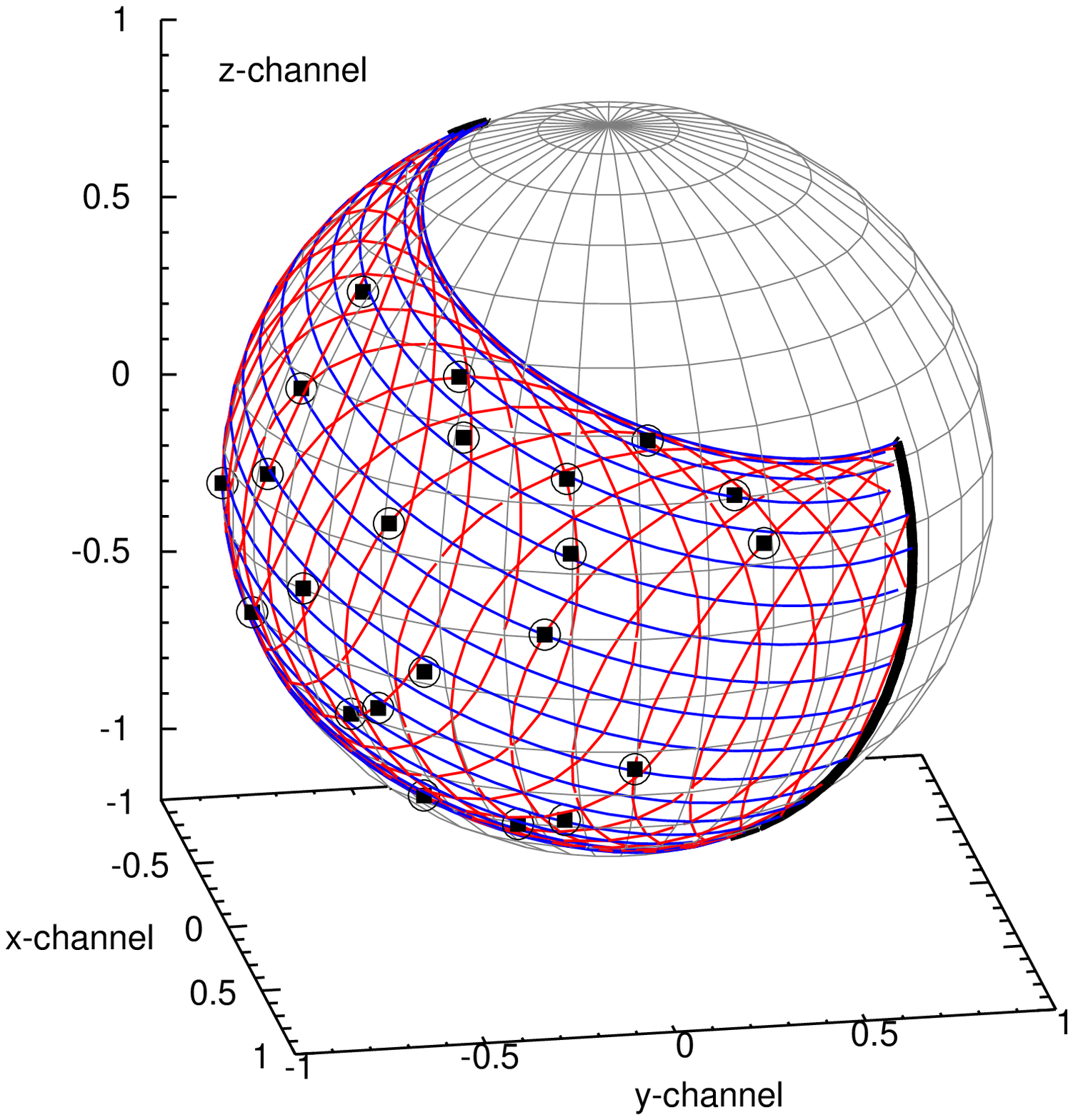}
\includegraphics[height=55mm]{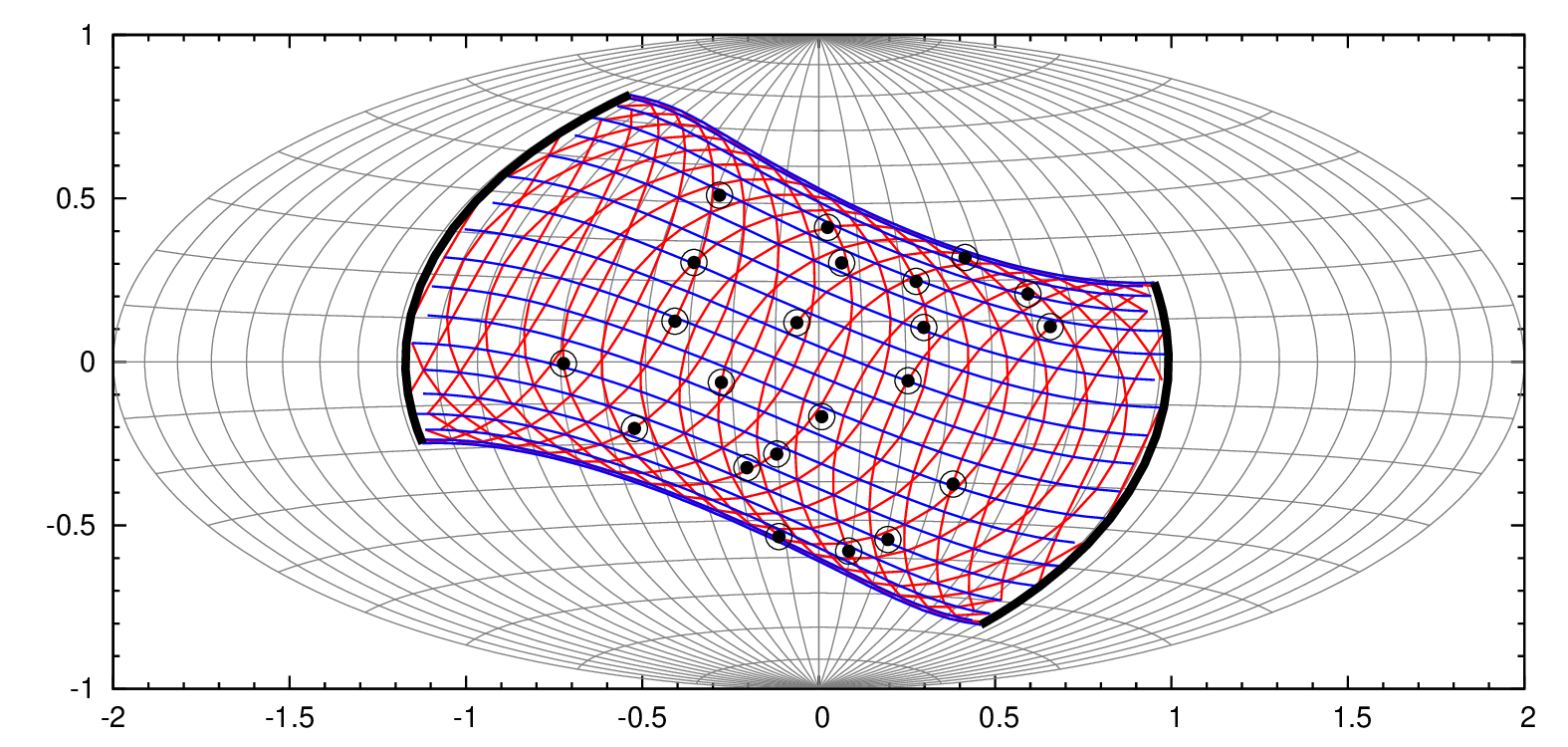}
\end{center}\vspace*{-4mm}
\caption{Contour lines of the accelerometer outputs as projected on 
the unit sphere. Red lines show curves for the same declination
(i.e. the hour angle is rotated) while blue lines show curves for
the same hour angle values (i.e. the declination axis is rotated).
Thick black lines mark the horizon. These maps show both configurations
of an equatorial mount, i.e. polar crossing is included. 
The left panel shows these contours on a sphere while the right panel
is plotted by involving an Aitoff projection (otherwise, the two panels
are equivalent). The black encircled
plots mark the points where the attitude calibration procedure has been
performed (see text for further details).}
\label{fig:isolines}
\end{figure*}

\subsection{Using the accelerometer as an attitude sensor}
\label{sec:accattitude}

\noindent
Let us suppose that an ideal static accelerometer is placed horizontally. 
In this case, its output vector is $\mathbf{a}=(x,y,z)=(0,0,g_0)$, where $g_0$ is magnitude
of the local gravity. It can be considered that if
the \emph{active} transformation $\mathbf{R}$ is applied on the 
accelerometer package, then the sensed vector is 
\begin{equation}
\mathbf{a}=-\mathbf{g}\cdot\mathbf{R}.\label{eq:accelerooutput}
\end{equation} 
Here $\mathbf{g}=(0,0,-g_0)$ is the
gravitational acceleration vector in the static (external) reference
frame defined by the $\mathbf{x}_0$, $\mathbf{y}_0$ and $\mathbf{z}_0$
axes. The active nature of the transformation means the following. Let 
us denote the reference frame fixed to the accelerometer package by 
the $\mathbf{x}$, $\mathbf{y}$ and $\mathbf{z}$ axes. Using these
notations, the active transformation means that 
\begin{eqnarray}
\mathbf{x} & = & \mathbf{R}\cdot\mathbf{x}_0, \\
\mathbf{y} & = & \mathbf{R}\cdot\mathbf{y}_0, \\
\mathbf{z} & = & \mathbf{R}\cdot\mathbf{z}_0.
\end{eqnarray}
In Fig.~\ref{fig:coord} we display the definitions of these vectors. 
In the following, we expect that the calibration procedure described in
Sec.~\ref{sec:calibration} has carefully been performed, thus the 
accelerometer output can be interpreted (within the accuracy of the 
calibration fit RMS) as an output of an ideal accelerometer.
One must note that the accelerometer output vector is computed 
as a \emph{post-multiplication} of the vector $-\mathbf{g}$ by the 
active rotation matrix $\mathbf{R}$. If $g_0$ is unity, then
the components of the output vector $\mathbf{a}=(a_1,a_2,a_3)\equiv(x'',y'',z'')$
are $a_1=R_{31}$, $a_2=R_{32}$ and $a_3=R_{33}$. Alternatively, one can
use the form
\begin{equation}
\mathbf{a}=\mathbf{R}^{\rm T}\cdot(-\mathbf{g}),
\end{equation}
where $(\cdot)^{\rm T}$ denotes matrix transposition. 

In the following, we investigate how the matrix $\mathbf{R}$ is computed
if the telescope parameters (most notably, the geographic latitude)
and the position of the axes are known.

\subsection{A simple isotropic pointing model}
\label{sec:isotropicpointing}

\noindent
As it was introduced above, the current goal is to compute the
transformation $\mathbf{R}$ that actively transform the accelerometer
from its ``rest'' position into the actual point where the telescope points.
Now we consider only an accelerometer that is mounted on the telescope
tube (referred as \#2 earlier above) since the role of the other one
on the hour axis is less crucial and the related computation is much 
simpler. Assuming an ideal construction, this matrix of $\mathbf{R}$
is computed as 
\begin{equation}
\mathbf{R}=\mathbf{G}\cdot(\mathbf{P}_{\rm t}\cdot\mathbf{P}_{\rm d})\cdot\mathbf{A}.\label{eq:acctrans}
\end{equation}
Here, $\mathbf{A}$ is the transformation that ``glues'' the accelerometer
chip to the tube, $\mathbf{P}_{\rm d}$ and $\mathbf{P}_{\rm t}$
are the transformations that rotates the telescope axes (declination
and hour axis, respectively) and $\mathbf{G}$ what ``installs'' the telescope
hour axis to its proper place on the ground. If we denote the
actual hour angle and declination values by $\tau$ and $\delta$, respectively
while the geographic longitude is $\varphi$, then these matrices are written as
\begin{eqnarray}
\mathbf{G} & = & \begin{pmatrix}\sin\varphi & 0 & -\cos\varphi\\0&1&0\\\cos\varphi&0&\sin\varphi\end{pmatrix}, \\
\mathbf{P}_{\rm t} & = & \begin{pmatrix}\cos\tau&\sin\tau&0\\-\sin\tau&\cos\tau&0\\0&0&1\end{pmatrix}, \\
\mathbf{P}_{\rm d} & = & \begin{pmatrix}\cos\delta&0&-\sin\delta\\0&1&0\\\sin\delta&0&\cos\delta\end{pmatrix},
\end{eqnarray}
while $\mathbf{A}$ depends on how we mounted the sensor to the tube. 
Throughout these computations, the reference axes $\mathbf{x}_0$, 
$\mathbf{y}_0$ and $\mathbf{z}_0$ point towards south, east and to the
zenith, respectively -- and hence form a right-hand coordinate system. 
In the case of an improper, but still isotropic alignment of the telescope,
the product $\mathbf{P}:=\mathbf{P}_{\rm t}\cdot\mathbf{P}_{\rm d}$ 
is written in the form
\begin{equation}
\mathbf{P}=\mathbf{H}\cdot\mathbf{P}_{\rm t}\cdot\mathbf{X}\cdot\mathbf{P}_{\rm d}\cdot\mathbf{T}.\label{eq:pointisotropic}
\end{equation}
Here the matrices $\mathbf{H}$, $\mathbf{X}$ and $\mathbf{T}$ encodes
the various misalignments, including polar misalignment, encoder zero points,
cross axis deflection, optical axis misalignment 
\citep[see e.g.][for a more detailed description of these deviations]{spillar1993}.
Ideally, all of transformations $\mathbf{H}$, $\mathbf{X}$ and $\mathbf{T}$
are unity. If these deflections are small, then a first-order expansion 
can be applied using the exponential form of ${\rm SO}(3)$ transformations:
\begin{eqnarray}
\mathbf{H} & = & \exp\begin{pmatrix}0&-c&b\\c&0&-a\\-b&a&0\end{pmatrix}
\approx\begin{pmatrix}1&-c&b\\c&1&-a\\-b&a&1\end{pmatrix}\\
\mathbf{X} & = & \exp\begin{pmatrix}0&-f&e\\f&0&-d\\-e&d&0\end{pmatrix}
\approx\begin{pmatrix}1&-f&e\\f&1&-d\\-e&d&1\end{pmatrix}\\
\mathbf{T} & = & \exp\begin{pmatrix}0&-i&h\\i&0&-g\\-h&g&0\end{pmatrix}
\approx\begin{pmatrix}1&-i&h\\g&1&-g\\-h&g&1\end{pmatrix}
\end{eqnarray}
For our purposes, ``small'' means that the second-order terms
are negligible compared to the pointing residual $\sigma_{\rm pointing}$.
This is limited now by the accuracy of the accelerometers, 
i.e. $\sigma_{\rm pointing}\approx\sigma_{\rm accelerometer}$. 
In other words, this condition is equivalent to 
\begin{equation}
a^2+b^2+c^2, d^2+e^2+f^2, g^2+h^2+i^2 \lesssim \sigma_{\rm pointing}.
\end{equation}
\def\cp{\ensuremath{c_{\rm p}}}\def\sp{\ensuremath{s_{\rm p}}}%
\def\ct{\ensuremath{c_{\rm t}}}\def\st{\ensuremath{s_{\rm t}}}%
\def\cd{\ensuremath{c_{\rm d}}}\def\sd{\ensuremath{s_{\rm d}}}%
It can be shown that the first-order series expansion 
of Eq.~(\ref{eq:pointisotropic}) depends only on the sums $c':=c+f$ and $e':=e+h$. 
For completeness, here we give the full expansion of 
this equation up to the first order. In the formula 
presented below, $\ct$, $\st$, $\cd$ and $\sd$ denotes
$\cos\tau$, $\sin\tau$, $\cos\delta$ and $\sin\delta$, respectively:
\begin{eqnarray}
\mathbf{P} & \approx & 
  \begin{pmatrix}\ct\cd&\st&-\ct\sd\\-\st\cd&\ct&\st\sd\\\sd&0&\cd\end{pmatrix}
+a\begin{pmatrix}0&0&0\\-\sd&0&-\cd\\-\st\cd&\ct&\st\sd\end{pmatrix}\label{eq:pexpand}\\
& &	+b\begin{pmatrix}\sd&0&\cd\\0&0&0\\-\ct\cd&-\st&\ct\sd\end{pmatrix}
	+c'\begin{pmatrix}\st\cd&-\ct&-\st\sd\\\ct\cd&\st&-\ct\sd\\0&0&0\end{pmatrix}+\nonumber \\
& & 	+d\begin{pmatrix}-\st\sd&0&-\st\cd\\-\ct\sd&0&-\ct\cd\\0&1&0\end{pmatrix}
	+e'\begin{pmatrix}\ct\sd&0&\ct\cd\\-\st\sd&0&-\st\cd\\-\cd&0&\sd\end{pmatrix}+\nonumber \\
& & 	+g\begin{pmatrix}0&-\ct\sd&-\st\\0&\st\sd&-\ct\\0&\cd&0\end{pmatrix}
	+i\begin{pmatrix}\st&-\ct\cd&0\\\ct&\st\cd&0\\0&-\sd&0\end{pmatrix}.\nonumber
\end{eqnarray}
If the transformation $\mathbf{A}$ is not accurately known, then
the respective corrections will appear in $\mathbf{T}$ via the
parameters $g$, $e'=e+h$ and $i$. This is due to the fact that in the final 
form of transformation Eq.~(\ref{eq:acctrans}), only the product
$\mathbf{T}\cdot\mathbf{A}$ appears. 
In the following, we proceed with the determination of the
pointing model parameters ($a$, $b$, $c'$, \dots). 

\subsection{Attitude calibration}
\label{sec:attcalib}

\noindent
In order to evaluate Eq.~(\ref{eq:accelerooutput}), we have to know not only
the pointing parameters and transformations but the accelerometer attitude 
$\mathbf{A}$ with respect to the telescope tube. As we noted earlier,
in our experiment we mounted the tube unit in a somehow random attitude
due to the limited mounting possibilities
(see Fig.~\ref{fig:accelerometers}\,b). However, the corresponding 
transformation can easily be estimated by combining some rotations
whose product yields the desired attitude. Our findings for this
attitude was
\begin{equation}
\mathbf{A}=\begin{pmatrix}
+0.6307 & -0.7759 & -0.0135 \\
-0.3365 & -0.2577 & -0.9057 \\
+0.6993 & +0.5758 & -0.4237
\end{pmatrix}.\label{eq:schmidtattitude}
\end{equation}
Before comparing the expected accelerometer outputs with the measured ones,
we have to multiply Eq.~(\ref{eq:pexpand}) by $-\mathbf{g}\cdot\mathbf{G}$
from the left and by $\mathbf{A}$ from the right. However, it is easier
to multiply the accelerometer outputs $\mathbf{a}$ by the transpose (inverse)
of $\mathbf{A}$, thus our constraint will be the relation
\begin{equation}
-\mathbf{g}\cdot\mathbf{G}\cdot(\mathbf{P})=\mathbf{a}\cdot\mathbf{A}^{\rm T}.\label{eq:accpointbasic}
\end{equation}
If a series of $\mathbf{a}_k$ values are given with the corresponding
$\tau_k$, $\delta_k$ values, one should minimize the merit function
\begin{equation}
\chi^2=\sum_k \left(-\mathbf{g}\cdot\mathbf{G}\cdot\mathbf{P}_k-\mathbf{a}_k\cdot\mathbf{A}^{\rm T}\right)^2\label{eq:chiaccelero}
\end{equation}
in order to find the best-fit values of the pointing model parameters
$a$, $b$, $c'$, $d$, etc. Since $\mathbf{g}=(0,0,-1)$ (considering unity
local gravitational acceleration), the components of the 
vector $-\mathbf{g}\cdot\mathbf{G}$ are going to be
\begin{equation}
-\mathbf{g}\cdot\mathbf{G}=\begin{pmatrix}\cos\varphi\\0\\\sin\varphi\end{pmatrix}.
\end{equation}
By multiplying Eq.~(\ref{eq:pexpand}) with this vector from the left, 
we got the expansion
\begin{eqnarray}
& & -\mathbf{g}\cdot\mathbf{G}\cdot\mathbf{P} =
\begin{pmatrix}\cp\ct\cd+\sp\sd\\\cp\st\\-\cp\ct\sd+\sp\cd\end{pmatrix}+
	a\begin{pmatrix}-\sp\st\cd\\\sp\ct\\\sp\st\sd\end{pmatrix}+\label{eq:accexpand} \\
& & +	b\begin{pmatrix}\cp\sd-\sp\ct\cd\\-\sp\st\\\cp\cd+\sp\ct\sd\end{pmatrix}+
	c'\begin{pmatrix}\cp\st\cd\\-\cp\ct\\-\cp\st\sd\end{pmatrix}+
	d\begin{pmatrix}-\cp\st\sd\\\sp\\-\cp\st\cd\end{pmatrix}+\nonumber\\
& & +	e'\begin{pmatrix}\cp\ct\sd-\sp\cd\\0\\\cp\ct\cd+\sp\sd\end{pmatrix}+
	g\begin{pmatrix}0\\-\cp\ct\sd+\sp\cd\\-\cp\st\end{pmatrix}+\nonumber\\
& & 	+i\begin{pmatrix}\cp\st\\-\cp\ct\cd-\sp\sd\\0\end{pmatrix}.\nonumber
\end{eqnarray}
Here $\cp=\cos\varphi$ and $\sp=\sin\varphi$. It can be computed that
the three vectors whose coefficients are $a$, $b$ and $c'$ are linearly
dependent for arbitrary values of $\varphi$, $\tau$ and $\delta$. This
property is implied by the fact that the accelerometer output is invariant
for the rotations around the $\mathbf{z}_0\pm$ axis. Due to the several
subsequently applied transformations needed to compute the final $\mathbf{a}$
vector, this invariance is not obvious at the first glance and appears
indirectly via this lost of linear independence. All of the other vectors
appearing in the equation above are independent in this sense. For 
simplicity, in the following we cancel the term corresponding to $c'$. 

Now it is straightforward to perform the minimization of 
Eq.~(\ref{eq:chiaccelero}). In order to test the above computations and
hence estimate the real-life accuracy of the sensors, we took 23 images
by the Schmidt telescope to gather sufficient (and unbiased) information
about the pointing of the accelerometers. The basic steps of image 
reduction are performed with the FITSH package \citep{pal2012} while 
astrometry is performed both by this package 
\citep[using the USNO-B catalog of][as reference]{monet2003}
and the online version of the Astrometry.net project \citep{lang2010}. 
The corresponding values for the hour angle ($\tau$) and 
declination ($\delta$)
that are needed in the expansion of Eq.~(\ref{eq:accexpand}) could be taken
from both by the astrometric solutions and from the rotary encoders mounted
inside the driving mechanisms of the telescope mount. First, J2000
centroids must be converted to first equatorial system for the epoch
of image acquisition (by taking into account precession, nutation, aberration
and refraction). For this purpose, we involved the algorithms
provided by \cite{meeus1998}. The linear regression is performed 
by using the algorithms of \cite{press2002} and the implementation provided
by the FITSH utility \texttt{lfit} \citep{pal2012}. The results of the fit are
\begin{eqnarray}
a  & = & -0.00091 \pm 0.00017 \\
b  & = & +0.00019 \pm 0.00010 \\ 
d  & = & -0.00011 \pm 0.00021 \\ 
e' & = & +0.00975 \pm 0.00011 \\ 
g  & = & +0.00082 \pm 0.00013 \\ 
i  & = & -0.00070 \pm 0.00025,
\end{eqnarray}
while the fit residual is $0.00025$. This value is equivalent to
$0.0143^\circ=0.86^{\prime}=52^{\prime\prime}$. Therefore, we can safely
conclude that accelerometers provide the sub-arcminute accuracy
as the part of a real TCS. In addition, the timespan between the 
calibration of the sensors and this attitude fit was approximately two months.
We note, there are other noise sources that are present in accelerometer
systems and not quantified by any of our calibration steps. These 
include the effect of mechanical vibrations of the telescope system as well as 
the Allan variance presented in the output of MEMS accelerometers. However,
the gross yield of these are also included in the final fit residual. 
The magnitude of long-term systematic variations can be characterized 
by repeating the attitude calibration over longer timespan. In addition,
the telescope itself can be utilized as a two-axis device
in order to estimate other sources for the temporal variations in,
at least, the affine part of the calibration procedure. 

\subsection{Extraction of pointing information}
\label{sec:extractionofpointing}

\noindent
Expecting an accelerometer to be the \emph{primary} absolute
pointing encoder of a telescope system, one can be curious how the
pointing (i.e. the $\tau$ and $\delta$ angles) can be recovered from
the accelerometer outputs. In Fig.~\ref{fig:isolines} we plotted the 
contour lines (more specifically, the isolines) of an equatorial telescope
located on the temperate geographical latitude $\varphi=47.5^\circ$. 
In these plots, the attitude $\mathbf{A}$ of the accelerometer with 
respect to the telescope is the same what it was in our experiments
(see Eq.~\ref{eq:schmidtattitude}). The topology of the contour lines
shows the ambiguity of the accelerometer outputs: there are 
positions corresponding to different $(\tau,\delta)$ values which yield
the similar $\mathbf{a}$ output. However, it can be considered
that this ambiguity is bimodal and can safely be resolved once 
the value of $\tau$ is known. Therefore, an accelerometer-based TCS
should employ \emph{two} such units: one is mounted on the polar axis
while the other one is fixed to the tube itself -- and even a rough value for
the $\tau$ is sufficient to resolve the bimodality. 

In practice, one has to invert Eq.~(\ref{eq:accpointbasic}) by substituting
the expression of Eq.~(\ref{eq:accexpand}) where the latter one is
a function of $\tau$ and $\delta$. Due to the first-order expansion, it can be
performed in an iterative way. First, one solve the equation
\begin{equation}
\begin{pmatrix}
\cos\varphi\cos\tau\cos\delta+\sin\varphi\sin\delta \\
\cos\varphi\sin\tau \\
-\cos\varphi\cos\tau\sin\delta+\sin\varphi\cos\delta
\end{pmatrix}=\begin{pmatrix}a_x\\a_y\\a_z\end{pmatrix},
\end{equation}
for $(\tau,\delta)$ where $(a_x,a_y,a_z)$ are the components of 
the product $\mathbf{a}\cdot\mathbf{A}^{\rm T}$.
Then, the solution is substituted to the first-order terms (proportional
to $a$, $b$, $d$, \dots) and subtracted from $(a_x,a_y,a_z)$ and
the iteration is repeated until convergence. The solution
of the above equation is going to be
\begin{eqnarray}
\tau & = & 90^\circ\pm \mathrm{arc\,cos}\left(\frac{a_y}{\cos\varphi}\right), \label{eq:acctauinverse} \\
\delta & = & \mathrm{arg}\left(d_x,d_y\right),
\end{eqnarray}
where
\begin{eqnarray}
d_x & = & a_x\cos\varphi\cos\tau+a_z\sin\varphi, \\
d_y & = & a_x\sin\varphi-a_z\cos\varphi\cos\tau.
\end{eqnarray}
The bimodality in the hour angle, i.e. the sign in Eq.~(\ref{eq:acctauinverse})
can safely be figured out by using a secondary accelerometer. Once
$\tau$ is known accurately, the value for $\delta$ is unambiguous.

The characteristics of the isolines in Fig.~\ref{fig:isolines} are 
also prominent. Naively, one can expect that due to the limits implied by
the local horizon, only the half of the sphere is covered by accelerometer.
However, due to the finite angle between the horizon and the primary axis
of the telescope, further information is lost: only a stripe in the sphere
is covered (which is also cut in half due to the horizon). The area of
this partial stripe relative to the total surface is $\cos\varphi/2$. The 
northern or southern the telescope location, the smaller the covered area.
In the poles, equatorial mounts behave similarly as alt-azimuthal mounts
and hence accelerometers could not provide sufficient information for
the pointing attitude. 

Exploiting the relation for horizontal altitude $h$, it can be 
computed rather elegantly, viz.
\begin{equation}
\sin h=a_x.
\end{equation}
The above equation shows how accelerometer outputs can be interpreted if
this sensor is used as a horizontal limit switch. We note here, however,
that fast slewing and the implied centrifugal acceleration distort the 
output and it must therefore be quantified before such an application. 
The magnitude of this distortion depends on both the slewing speed
and the displacement of the sensors from the axes. Considering a normal
slewing speed of two degrees per second (i.e. $0.03\,{\rm rad}/{\rm s}$)
and characteristic instrument size of a few meters, this centrifugal 
acceleration is going to be in the range of $\lesssim 10^{-3}\,g$,
which is equivalent to few arcminutes. 

\section{Summary}
\label{sec:summary}

\noindent
In this paper we demonstrated how cheap MEMS accelerometers can accurately
be calibrated and involved as a part of a telescope control system. 
Our main conclusion can be summarized in a single number, namely the 
fit residual of the telescope pointing fit procedure. This value shows
us that the sub-arcminute RMS accuracy can safely be targeted. This
accuracy includes the accuracy of the standalone calibration, the 
compensation of effects due to the variations in the ambient temperature
as well as the compensation of telescope mount deflections. 

Considering the techniques of the implementation, exploiting such sensors
needs no changes in the existing electromechanical components of a telescope
system at all. The sensors are simply mounted on the respective mechanisms
(hour axis and tube) and the actual attitude of the mounting is also irrelevant.
This is a great advantage over the electromechanical feedback systems
widely employed in TCSs.  In addition, the $2\times 3$ channels are
redundant and very sensitive of unintentional tampering of these devices.
This property further increases the reliability of an autonomous and/or
remotely operated observatory. 

One of the further goals of ours is to exploit such an accurately calibrated 
set of accelerometers in unconventional mechanics like hexapods
\citep{chini2000,koch2009,pal2013}. In this case, accelerometers can 
be mounted onto the base and payload platform as well as on all of the
six, topologically identical legs. The information provided by the 
$8\times3$ accelerometer channels can be sufficient and redundant to 
recover the attitude and displacement (i.e. 6 degrees of freedom)
of the payload with respect to the base. Our findings for the accuracy
is comparable to the per-pixel resolution of wide-field hexapod-based
instruments \citep{pal2013,vida2014}. Hence, such systems might benefit
even more than high resolution instrumentation. 

\acknowledgments
This research is conducted as a part of the ``Fly's Eye'' project which 
is supported by the Hungarian Academy of
Sciences via the grant LP2012-31. Additional support
is also received via the OTKA grants K-109276 and K-104607. We
thank F. Schlaffer for aiding the fabrication of the accelerometer
enclosures. We also thank the help and quick responses of our colleagues, 
R.~Szak\'ats, K.~Vida, Gy.~Mez\H o, M.~R\'acz, L.~Moln\'ar and 
L.~D\"obrentei during the installation and calibration of the sensors.
We also thank the valuable comments and suggestions of the anonymous referee. 
In our project, we involved numerous free \& open source software, including 
gEDA (for schematics and PCB design), OpenSCAD (3D parametric designs),
FreeCAD (3D designs), CURA (3D slicing, GCODE generation and printing control) 
and AVR-GCC (for MCU programming).

{}

\end{document}